\documentclass[12pt,showpacs,pre]{revtex4}
\usepackage{amsfonts,amssymb,amsmath,graphicx}
\newcommand{\ud}{\mathrm{d}}
\newcommand{\ic}{\mathrm{i}}

\newcommand{\re}{\mathrm{ Re }\ }

\newcommand{\zZ}{\mathbb{Z}}
\newcommand{\sgn}{\mathrm{sgn}}

\begin{document}
\title{Fidelity for kicked atoms with gravity near a quantum resonance}
\author{R\'emy Dubertrand}
\affiliation{School of Mathematics, University of Bristol, University Walk,
  Clifton, Bristol BS8 1TW, United Kingdom}
\affiliation{Institut f\"ur theoretische Physik,
  Ruprecht-Karls-Universit\"at, Philosophenweg 19, 69120
  Heidelberg, Germany}
\author{Italo Guarneri}
\affiliation{Center for Nonlinear and Complex Systems, Universit\'a
  dell'Insubria, Via Valleggio 11, 22100 Como, Italy}
\author{Sandro Wimberger}
\affiliation{Institut f\"ur theoretische Physik,
  Ruprecht-Karls-Universit\"at, Philosophenweg 19, 69120
  Heidelberg, Germany}
\affiliation{Heidelberg Center of Quantum Dynamics, Philosophenweg 12, 69120
  Heidelberg}

\begin{abstract}
Kicked atoms under a constant Stark or gravity field are investigated for experimental
setups with cold and ultra cold atoms. The parametric stability of the
quantum dynamics is studied  using
the fidelity. In the case of a quantum resonance, it is shown that the behavior of the fidelity
depends on arithmetic properties of the gravity parameter.
Close to a quantum resonance, the long time asymptotics of the fidelity is studied by means of
a {\em pseudo-classical} approximation first introduced by Fishman {\em et al.} [J. Stat. Phys. {\bf 110}, 911 (2003)].
The long-time decay of fidelity arises from the tunneling out of pseudo-classical stable islands,
and a simple ansatz is proposed which satisfactorily reproduces the main features observed in numerical simulations.
\end{abstract}
\maketitle

\section{Introduction}

The stability of quantum evolution against parametric changes of the
quantum Hamiltonian is a subject
of wide theoretical and experimental interest. A widely used concept
is the fidelity introduced by Peres
\cite{peres}, and the closely related Loschmidt echo \cite{JP2001,Gorin2006}, which is  built as the interference
pattern between states which are obtained by propagating the same initial state
under Hamiltonians, say $\hat{H}_0$ and $\hat{H}$, which are slight
perturbations of each other. A standard definition of  the fidelity is:
\begin{equation}
  \label{defF}
  F(t)=\left| \langle \psi | e^{\ic \hat{H_0} t/\hbar} e^{-\ic \hat{H} t/\hbar} | \psi \rangle \right|^2 .
\end{equation}
The behavior of fidelity in time is known to display some universal properties that reflect
the underlying classical dynamics \cite{JP2001,Gorin2006}. Such properties have been mostly
explored for the case of systems, which are chaotic %either chaotic or integrable 
in the classical limit.
In this paper we will study the fidelity  in the  mixed phase space regime. For this the system
under consideration is the quantum kicked rotor \cite{Casati,chirikov,raizen}.

The main motivation for our analysis here is twofold. Firstly, there has been a growing interest over the
last decade in the dynamics of the quantum kicked rotor (and its variants) at and close to the so-called
quantum resonances \cite{QR},
theoretically (see, e.g, \cite{fishman2002,sandro2003,sheinman,kres,italo2006,sandro06,italo2008,abb,review}) as well as experimentally
(see, e.g., \cite{oberthaler,Gil2006,sadgrove,nist,sadgrove2007,harvard,Gil2010}). Secondly, only recently concepts
have been developed to actually access the fidelity in setups based on cold or ultracold quantum gases. The used
techniques range from interferometric methods in either internal atomic states \cite{exp} or in the center-of-mass motion
of the atoms \cite{harvard} to the time reversal of the dynamics by exploiting the properties of the quantum resonant motion
\cite{Gil2010,hoo2011}.

In this paper we study the quantum kicked rotor under the additional influence of a Stark or gravity field \cite{oberthaler,fishman2002}.
In Sect.~\ref{sectQKR} the Hamiltonian of one kicked atom is quickly reminded and the fidelity,
which is the main quantity studied here, is precisely defined for our system. We then report on a subtle dependence of the fidelity
on the arithmetic properties of the relevant parameters (Sect.~\ref{sectQR}),
a result which may be interesting for future precise measurements
of fundamental constants (see the discussions in refs. \cite{harvard,Gil2010,sack2011}).
Sect.~\ref{secteps} is devoted to the dynamics close to quantum resonance.
Based on the pseudo or $\epsilon-$classical formalism developed by Fishman {\em et al.}
\cite{fishman2002}, we explain the overall behavior of the fidelity using  $\epsilon-$classical phase space densities and
quantum tunneling rates from the stable resonance island to the surrounding chaotic sea
in phase space.  Some  technical details are found in appendices.

\section{The kicked rotor with gravity}
\label{sectQKR}

We are interested in the quantum dynamics of a particle moving in a line, periodically kicked in time, and
subject to a constant Stark or gravity field. It is described by the Hamiltonian in dimensionless
variables (such that $\hbar=1$) \cite{fishman2002}:
\begin{equation}
  \label{Hkickr}
  \hat{H}(\hat{x},\hat{p},t)\;=\;\frac{\hat{p}^2}{2}-\frac{\eta}{\tau}\; \hat{x}+ k\cos( \hat{x}) \sum_{\verb+t+\in \zZ}
\delta(t-\verb+t+\tau)\ .
\end{equation}
The kicking period is $\tau$,
the kicking strength is $k$, and  $\verb+t+$ is a discrete time variable that
counts the number of kicks.  The parameter $\eta$
yields the change in momentum produced by the constant field in one kicking period.
In the  accelerated frame of reference \cite{fishman2002},  the  potential
experienced by the particle is periodic in space and so  the quasi-momentum
$\beta$ is conserved by the evolution. With the chosen units, $\beta$  takes
all values between $0$ and $1$.  Using Bloch theory, the particle dynamics can then be identified with that of a
family of quantum  rotors, labelled by  the values of $\beta$.  For the $\beta$-rotor ({\it i.e.}, the
rotor in the family to which a given  value  $\beta$ of the quasi-momentum is
affixed) the evolution   from immediately after the $(\verb+t+-1)$-th
kick to immediately after the $\verb+t+$-th kick is described by the unitary propagator \cite{fishman2002}:
\begin{equation}
  \hat{\cal U}_{\beta,k,\eta}(\verb+t+)\;=\;e^{-\ic k \cos(\hat{\theta})}\; e^{-\ic \tau/2(\hat{\cal N}
 + \beta +\eta\verb+t+ +\eta/2)^2}\ ,
 \label{Flqt}
\end{equation}
and the evolution operator over the first $\verb+t+$ kicks is:
\begin{equation}
  \label{U_tot}
  \hat{\cal
     U}_{\beta,k,\eta}^{\verb+t+}\;\equiv \;\hat{\cal
     U}_{\beta,k,\eta}(\verb+t-1+)\; \hat{\cal U}_{\beta,k,\eta}(\verb+t-2+)
   \dots \hat{\cal U}_{\beta,k,\eta}(\verb+1+)\; \hat{\cal U}_{\beta,k,\eta}(\verb+0+)  \,.
\end{equation}
where $\hat{\cal N}$ is the momentum operator:
\begin{equation*}
  \hat{\cal N}=-\ic \frac{\ud}{\ud\theta}\ ,
\end{equation*}
with periodic boundary conditions. The time-dependent  Hamiltonian that generates the quantum evolution
corresponding to (\ref{U_tot}) is then:
\begin{equation}
  \label{Ham1}
  \hat{\cal H}({\hat{\cal N}},\hat{\theta},\beta,t)= \frac{1}{2}
  \left(\hat{\cal N} +
      \beta +\frac{\eta}{\tau} t\right)^2+ k\cos(\hat{\theta}) \sum_{\verb+t+\in\zZ}
    \delta(t-\verb+t+\tau)\ .
\end{equation}
We will study the fidelity that  measures the stability of the evolution (\ref{U_tot}) with respect to changes
of the parameter $k$. For a given $\beta$-rotor this fidelity is defined by:
\begin{equation}
  \label{fid_def1rot}
 F_\beta(k_1,k_2,\eta,\verb+t+) = \Big| \left< \hat{\cal
     U}_{\beta,k_1,\eta}^{\verb+t+} \psi \Big|\hat{\cal
     U}_{\beta,k_2,\eta}^{\verb+t+} \psi\right>\Big|^2\ ,
\end{equation}
Moreover, having in mind experimental situations with cold atoms \cite{exp,sandro2003,sadgrove,review,harvard}, we will also
consider the case when the initial state of the atomic cloud
is an incoherent mixture of plane waves with a distribution
$\rho(\beta)$ of the quasi-momentum. In this case
the fidelity is given  by \cite{sandro06}:
\begin{equation}
  \label{fid_ens_rot}
 F(k_1,k_2,\eta,\verb+t+) =\bigg|\int_0^1 \rho(\beta) \left< \hat{\cal
     U}_{\beta,k_1,\eta}^{\verb+t+} \psi \Big|\hat{\cal
     U}_{\beta,k_2,\eta}^{\verb+t+} \psi\right> \ud \beta \bigg|^2%  = \bigg|\int_0^1 \rho(\beta) \,.
% J_0(\Delta k |W_{\verb+t+}(\beta,\eta)|)\ud \beta \bigg|^2
\end{equation}

\section{Fidelity at a quantum resonance}
\label{sectQR}

In the gravity free case ($\eta=0$), Eq. (\ref{Flqt}) describes the standard
Kicked-Rotor (KR) dynamics, and so-called KR resonances \cite{QR,kres-1,kres} occur
whenever $\tau$ is commensurate to $2\pi$. Then, for special values of
quasi-momentum $\beta$, the energy of the $\beta$-rotor  asymptotically
increases quadratically as $\verb+t+\to\infty$.  In the presence of gravity,
asymptotic quadratic growth of energy at certain values of $\beta$ is still
possible \cite{dana07}.  Here we study the behavior of fidelity in the
presence of gravity  and for  the case of a main KR resonance,  i.e., $\tau=2\pi
l$ (with integer $l$).
Denoting $\psi_{\beta}(\verb+t+)\equiv\hat{\cal U}_{\beta,k,\eta}^{\verb+t+}\psi_{\beta}(0)$,
one may explicitly compute \cite{fishman2002}:
  \begin{equation}
    \label{psi_mom}
 \left<\theta|\psi_\beta(\verb+t+)\right>=e^{-\ic\alpha(\beta,\eta,\verb+t+)}
 e^{-\ic k A(\theta,\beta,\eta,\verb+t+)}
\bigl<\theta- \verb+t+(2\beta+1)\pi l-\pi l \eta
  \verb+t+^2 \;\big |\;\psi_\beta(0)\bigr>\,
  \end{equation}
where  $\alpha(\beta,\eta,\verb+t+)$ is a global phase, and
\begin{equation}
  \label{psiF}
  A(\theta,\beta,\verb+t+)=\sum_{r=0}^{\verb+t+-1}\cos\bigl(\theta-(2\beta+1)\pi
    l\;r-2\pi l r \eta \verb+t+ +\pi l\eta r^2\bigr)\ .
\end{equation}
From now on we assume that the initial state is a plane wave:
$\left|\psi_{\beta}(0)\right>=|n_0\rangle$, i.e.
\begin{equation}
  \label{psi_init_1}
  \left<\theta|\psi_\beta(0)\right>=\left<\theta|n_0\right>=\frac{e^{\ic n_0\theta}}{\sqrt{2\pi}}\ .
\end{equation}
The fidelity is directly obtained from (\ref{psi_mom})
using the method described in \cite{sandro06}. First write
$A(\theta,\beta,\verb+t+)=\re (e^{\ic\theta} W_{\verb+t+})$ with
\begin{equation}
 W_{\verb+t+}\equiv W_{\verb+t+}(\eta,\beta)=\sum_{r=0}^{\verb+t+-1}
e^{-\ic \pi  l(2\beta+1)r}e^{-\ic 2\pi l r \eta \verb+t++\ic\pi l\eta r^2
}\ .\label{Wt_1}
\end{equation}
Then  from (\ref{psi_mom}) it follows that the wave function after the $\verb+t+$-th kick is given
 in momentum representation by:
\begin{eqnarray}
  \label{iterpsi2}
  \left<n\left|\;\hat{\cal U}_{\beta,k_1,\eta}^{\verb+t+}\;\right|\psi\right>\;&=&\; e^{n \ic\arg
    (W_{\verb+t+})} \frac{1}{2\pi}\int_0^{2\pi}e^{\ic( n_0-n)\theta}
e^{-\ic k |W_{\verb+t+}|\cos\theta}\; d\theta\nonumber\\
&=&\;e^{n \ic\arg
  (W_{\verb+t+})}(-\ic)^{n_0-n} J_{n_0-n}(k |W_{\verb+t+}|)\,
\end{eqnarray}
where $J_{n_0-n}(.)$ is the Bessel function of order $n_0-n$.  Using
this along with  the addition formula of Bessel functions, see
e.g. 7.15.(31) in \cite{bateman2}, one finds:
\begin{equation}
  \left< \hat{\cal U}_{\beta,k_1,\eta}^{\verb+t+}
\psi \Big|\hat{\cal U}_{\beta,k_2,\eta}^{\verb+t+} \psi\right>= J_0(\Delta k
|W_{\verb+t+}|)\ ,
\label{overlap1rot}
\end{equation}
where we defined the perturbation parameter $\Delta k=k_2-k_1$, and so
the fidelity of a single $\beta-$rotor is given by:
\begin{equation}
  \label{fid_1rot}
   F_\beta(k_1,k_2,\eta,\verb+t+)= \Big|
J_0(\Delta k |W_{\verb+t+}|)\Big|^2 \,.
\end{equation}

\subsection{Asymptotics of the fidelity for one single rotor}

From Eqs.~(\ref{iterpsi2}) and (\ref{fid_1rot}) it is clear that the long time
asymptotics of the wave-packet propagation, and of the fidelity as well,
are determined by the  behavior of  $|W_{\verb+t+}|$ as $\verb+t+\to\infty$. One may write:
\begin{equation}
   W_{\verb+t+}\;=\;e^{i\Phi(\eta,\beta,\verb+t+)}\;\mathfrak{W}(\beta,\eta,\verb+t+)\;,
\end{equation}
where $\Phi(\eta,\beta,\verb+t+)= - \pi l (2\beta+1+\eta\verb+t+)\verb+t+$, and
$\mathfrak{W}(\eta,\beta,\verb+t+)$ is a quadratic Weyl sum:
\begin{equation}
\label{qweyl}
\mathfrak{W}(\eta,\beta,\verb+t+)\;= \;\sum_{r=1}^{\verb+t+}
e^{\ic\pi l(2\beta+1)\;r}\;e^{\ic\pi l\eta\; r^2} \,.
\end{equation}
The asymptotic behavior of such sums as $\verb+t+\to\infty$ is known to
depend on the arithmetic nature of the number $\eta$, i.e., on whether it
is  rational or irrational, and in the latter case on its Diophantine
properties  \cite{berry}.  First of all, as the behavior of Weyl sums may be
quite erratic, we resort to the time-averaged fidelity defined by:
\begin{equation}
  \label{Fid_T_av}
  \left<F_\beta(k_1,k_2,\eta,\verb+t+)\right>_T\equiv\frac{1}{T}
  \sum_{\verb+t+=0}^{T-1} F_\beta(k_1,k_2,\eta,\verb+t+), \quad T\gg 1 \,,
\end{equation}
which has a smoother dependence on time than the original fidelity.

The easiest case is when $\eta$ is rational: $\eta=p/q$, with $p$ and
$q$ mutually prime integers. In that case, setting
$r=2jq+\nu$ in the sum (\ref{qweyl}) with $j$ a non negative integer
and $0\leq \nu\leq 2q-1$, the sum may be rewritten in the form:
$$
 \mathfrak{W}(\eta,\beta,\verb+t+)\;=\;C(\eta,\beta,\verb+t+)\;B(\eta,\beta,\verb+t+)\;,
 $$
where
\begin{eqnarray}
C(\eta,\beta,\verb+t+)\;&=&\;\sum\limits_{j=0}^{[\verb+t+,2q]}e^{-4i\pi l\beta jq}\;,\nonumber\\
B(\eta,\beta,\verb+t+)\;&=&\;\sum\limits_{\nu=0}^{\{\verb+t+,2q\}}e^{i\pi l(2\beta+1)\nu}\;e^{i\pi l\nu^2p/q}\;,
\end{eqnarray}
having denoted by $[\verb+t+,2q]$ the integer part of $\verb+t+/(2q)$ and $\{\verb+t+,2q\}=\verb+t+$ mod$(2q)$.
The factor $B(\eta,\beta,\verb+t+)$ is a periodic function of
$\verb+t+$ with period $2q$. Explicit calculation of the sum on the
right hand side of the 1st equation shows that $C(\eta,\beta,\verb+t+)$ is:
\begin{itemize}
\item quasi-periodic for $\beta$ irrational,
\item periodic  for $\beta$ rational and $2\beta q$ non-integer,
\item linear, i.e. $C(\eta,\beta,\verb+t+)=\verb+t+$, when $2\beta q$ is an integer.
\end{itemize}
Such facts have the following implications on wave packet dynamics on
the one hand and on the behavior of the fidelity on the other
hand. Two cases have to be distinguished, according to whether $2\beta
q$ is integer, or not. In the former case, a quantum resonance
occurs. Indeed,  using  that $|W_{\verb+t+}|$ is equal to $\verb+t+$
times a periodic function of $\verb+t+$,  from Eq.~(\ref{iterpsi2}) and from
 the well-known asymptotics of the Bessel functions at large argument
 and fixed order, see e.g. 7.13.1 (3) in \cite{bateman2}:
\begin{equation}
\label{bessasy}
  J_n(x)\sim\sqrt{\frac{2}{\pi x}}\cos\left(x-\frac{n\pi}{2}-\frac{\pi}{4}\right), \quad x\to\infty \,,
\end{equation}
we find that the probability in the $n$-th momentum eigenstate decays in time
like $1/\verb+t+$. Hence the wave packet spreads
linearly in time in momentum space, and the energy quadratically
increases. Instead, if $2\beta q$ is not an integer then Eq.~(\ref{iterpsi2})
shows that  the amplitude of the
evolving  wave function in any momentum eigenstate  oscillates
quasi-periodically in time, so no unbounded spreading in momentum space occurs.

\begin{figure}[!ht]
\begin{minipage}{.49\linewidth}
  \begin{center}
    \includegraphics[width=\textwidth]{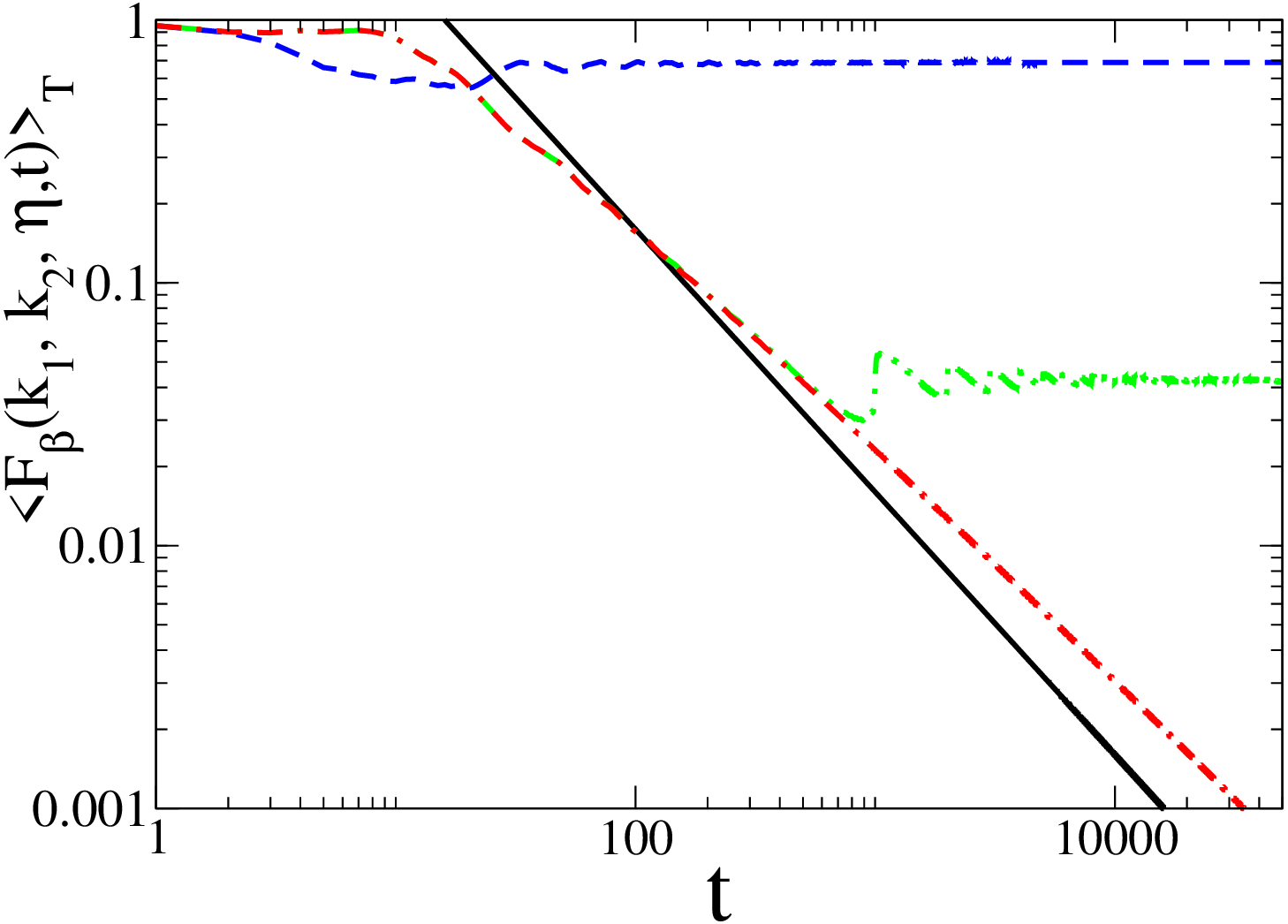}
\vspace{0.3cm}
    a)
  \end{center}
\end{minipage}
\begin{minipage}{.49\linewidth}
  \begin{center}
    \includegraphics[width=\textwidth]{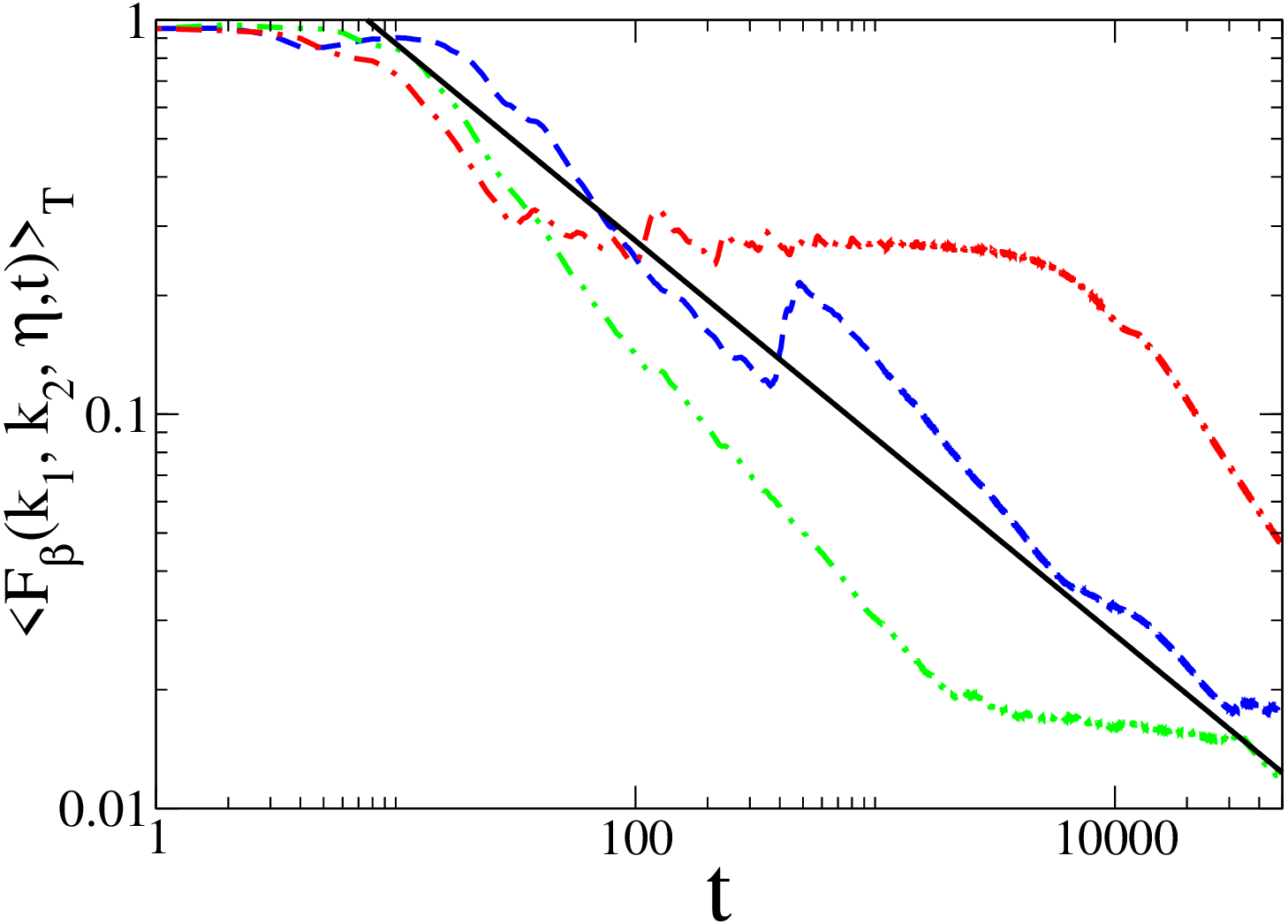}

\vspace{0.3cm}
    b)
  \end{center}
\end{minipage}
\caption{(Color online) Fidelity for the kicked rotor with gravity at the quantum
  resonance $\tau=2\pi$ for $k_1=0.8\pi$ and $k_2=0.7\pi$. The initial state is a plane wave with zero
  momentum $n_0=0$. Several choices of $\eta$ and $\beta$ are
  shown:  a)
  $\eta=0.1$. Full black line: const.$/{t}$. Blue dashed line: $\beta=0.23$. Green
  double dot-dashed line: $\beta=0.499$. Red dot-dashed line:
  $\beta=0.5$. b) Full black line: const.$/\sqrt{t}$. Blue dashed line:
  $\eta=(\sqrt{5}-1)/2$ (the Golden Ratio), $\beta=0.23$. Green double dot-dashed line: $\eta=\pi$,
  $\beta=0.23$. Red dot-dashed line: $\eta=\pi$, $\beta=0.5$.}
\label{Fid_T_av_1rot}
\end{figure}

A similar reasoning based on  Eqs.~(\ref{fid_1rot}) and (\ref{Fid_T_av}) shows
that, for all nonresonant values of $\beta$,   the time-averaged fidelity
saturates to a nonzero value in the limit $T\to\infty$
(Fig.~\ref{Fid_T_av_1rot}a). Instead, at resonance ($2\beta q=$integer) from
Eqs.~(\ref{fid_1rot}) , (\ref{bessasy}) (with $n=0$) and  (\ref{Fid_T_av}), the
fidelity is seen to asymptotically decay to zero like
$\log(T)/T$ (case $\beta=0.5 $ in Fig.~\ref{Fid_T_av_1rot}a).
\\ The case of irrational $\eta$ is much more difficult, as the behavior of
Gauss
sums crucially depends on the Diophantine properties of $\eta$
\cite{berry}. Here we limit ourselves to a heuristic analysis.
For strongly irrational $\eta$ one may naively picture $W_{\verb+t+}$ as a
sort of random walk in the complex plane, suggesting that $|W_{\verb+t+}|$ grows 
like $\sqrt{\verb+t+}$ in some average sense. Thanks to (\ref{fid_1rot}) and
(\ref{bessasy}) an asymptotic decay of the average fidelity like
$1/\sqrt{\verb+t+}$ is then expected. This is roughly numerically confirmed
for the case when $\eta$ is equal to the golden ratio in Fig.~\ref{Fid_T_av_1rot}b. However the
actual decay displays strong fluctuations because it depends on
the continued fraction expansion of $\eta$, notably large partial quotients in
the latter may cause fidelity to behave as in cases of rational $\eta$ over
significant time scales, e.g., for $\eta=\pi=[3,7,15,1,292,1,1,\dots]$
in Fig.~\ref{Fid_T_av_1rot}b.

\begin{figure}[!ht]
\centering
%\vspace{1cm}
%\includegraphics[width=.5\textwidth]{Fid_T_average_QR_ens.eps}
\includegraphics[width=.5\textwidth]{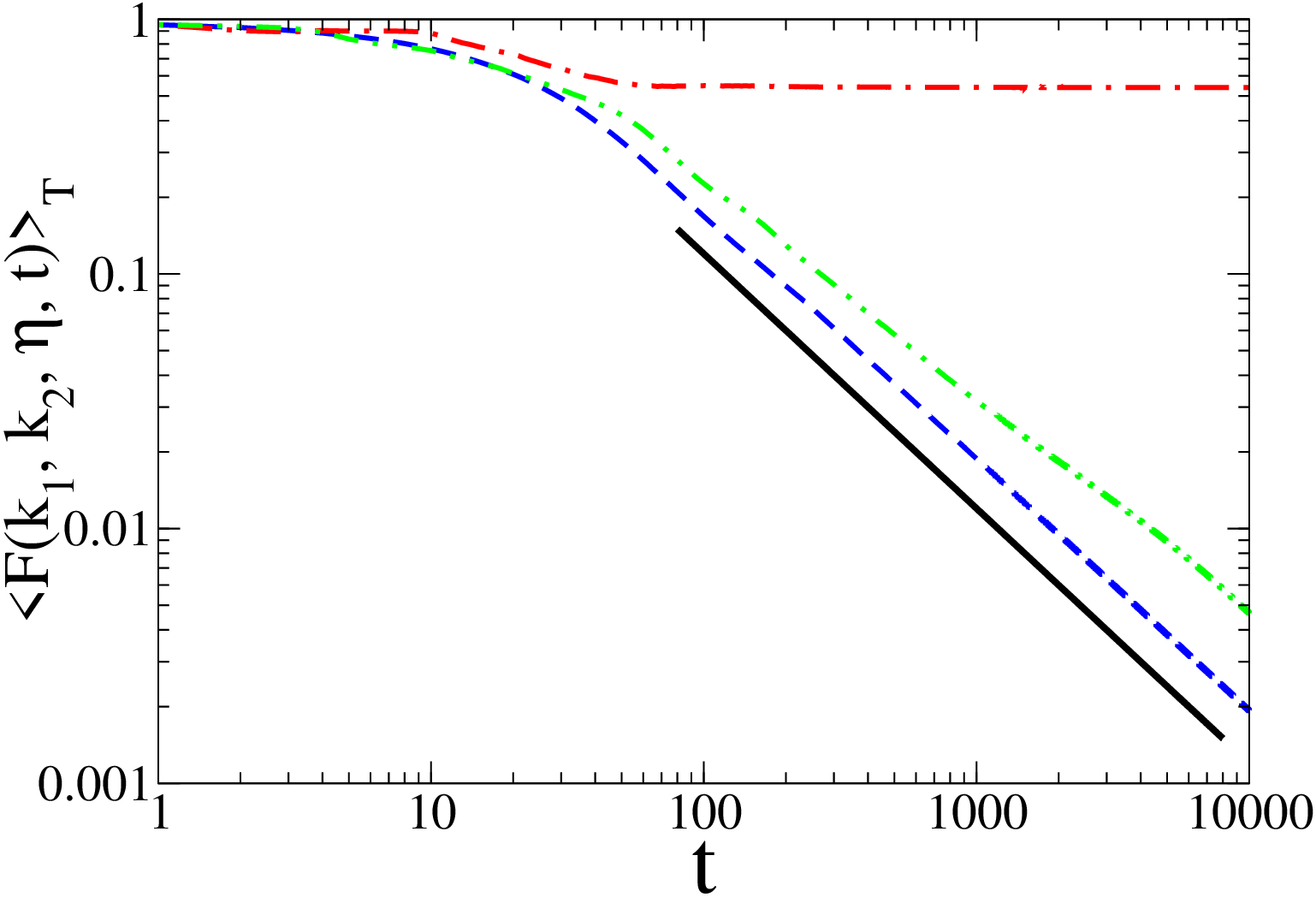}
\caption{(Color online) Fidelity for an ensemble of kicked rotors with gravity
  at $\tau=2\pi$ for $k_1=0.8\pi$ and $k_2=0.7\pi$. 
  The initial state is a plane wave with $n_0=0$. Several choices of $\eta$ are
  shown. The integral over $\beta$ in Eq.~(\ref{fid_ens_rot}) is computed via a Riemannian sum
  with $5000$ randomly chosen $\beta$s. Full black line: const.$/t$. Blue
  dashed line: $\eta=$Golden Ratio. Green double dot-dashed line: $\eta=\pi$. Red
  dot-dashed line: $\eta=0.1$.}
\label{ens_rotQR}
\end{figure}

%\subsection{Asymptotics of the fidelity for an ensemble of rotors.}
In order to mimic the experimental setups based on cold atoms
\cite{exp,sandro2003,sandro06,sadgrove,harvard,review},
we consider the case when the initial state of the kicked atoms is an incoherent mixture of plane waves
with  a uniform density of $\beta$: $\rho(\beta)=1$. The expression (\ref{fid_ens_rot}) is computed
as an average over a large number of randomly chosen values of $\beta$. The
result  does not vary significantly when the number of
values of $\beta$ exceeds a few thousands. We  observe a sharp difference in
the asymptotic regime depending on whether $\eta$ is rational or not, see
Fig.~\ref{ens_rotQR}. We again show the time-averaged
quantity:
\begin{equation}
  \label{Fid_ens_T_av}
  \left<F(k_1,k_2,\eta, \verb+t+ )\right>_T \equiv \frac{1}{T}
  \sum_{\verb+t+ = 0}^{T-1} F(k_1,k_2,\eta, \verb+t+ ) \,.
\end{equation}
For rational values of $\eta$, the fidelity is observed to saturate towards a finite
value. This is not surprising, because this is precisely the expected behavior for all
values of $\beta$ in a set of full measure. On the contrary, the fidelity decays like $1/T$ for
irrational $\eta$, see Fig.~\ref{ens_rotQR}. This is roughly explained noting
that, besides the decaying prefactor $\sim |W_{\verb+t+}|^{-1/2}$ in the asymptotics (\ref{bessasy}) (with $n=0$),
one more mechanism of decay is introduced by ensemble averaging, which
affects  the rapidly oscillating part of the Bessel function.
The observation of the fine details of the number theoretical properties of $\eta$ is
certainly of high interest for precision measurements
\cite{Gil2010,harvard,hoo2011,sack2011} but, at the same time, a big challenge to
experimental resolution in current setups.

\section{Fidelity close to a quantum resonance.}
\label{secteps}

\subsection{Reminder of the $\epsilon$-semiclassics. Fixed points.}
When the kicking period is close to a quantum resonance,
\begin{equation}
  \label{tau}
  \tau=2\pi l + \epsilon,\qquad l \textrm{ integer }, |\epsilon|\ll 1,
\end{equation}
we will implement a technique of quasi-classical approximation
 originally described in
\cite{fishman2002} and therein termed ``$\epsilon$-classical approximation''.
Introducing a rescaled kick strength  $\tilde{k}=|\epsilon| k$, and a new momentum operator
\begin{equation}
  \hat{I}=|\epsilon|\hat{\cal N}=-\ic |\epsilon| \frac{\ud}{\ud\theta}\,,
\label{scaledmom}
\end{equation}
the propagator (\ref{Flqt}) may be rewritten as:
\begin{equation}
  \label{Flqt2}
  \hat{\cal U}_{\beta,k,\eta}(\verb+t+)=\exp\left(-\frac{\ic}{|\epsilon|} \tilde{k}
    \cos(\hat{\theta})\right) \exp\left( -\frac{\ic}{|\epsilon|}{\hat{\cal H}}_{\beta}(\hat{I},\verb+t+)\right)\ ,
\end{equation}
where:
\begin{equation}
  \label{effH}
  \hat{\cal H}_{\beta}(\hat{I},\verb+t+)\;=\;
  \frac12\sgn(\epsilon){\hat{I}^2}\;+\;\hat{I}\bigl[\pi l
  + \tau (\beta+\eta\verb+t++\frac{\eta}2) \bigr].
\end{equation}
The small parameter $\epsilon$ plays the formal role of a Planck constant in (\ref{Flqt2}), so,
close to a quantum resonance, the quantum dynamics mirrors an ``$\epsilon$-classical''
dynamics, which is
 immediately inferred from (\ref{Flqt2}) and (\ref{effH}). After changing
 the $\epsilon$-classical momentum variable from $I$ to $J$ given by:
\begin{equation}
  \label{ItoJ}
  % \left\{
  %   \begin{array}{rcl}
      {J}={I}+\sgn(\epsilon)[\pi l+\tau\beta+\tau\eta\verb+t++\tau\eta/2]\ ,%\\
      % \end{array}\right. \ .
\end{equation}
the $\epsilon$-classical dynamics is described by the following map that relates
the variables $J$ and $\theta$ from immediately after the $\verb+t+$-th kick to immediately after the $(\verb+t++1)$-th one:
\begin{equation}
  \label{classical}
  \left\{
    \begin{array}{rcl}
      J_{\verb+t++1}&=&J_{\verb+t+}\;+\;\tilde{k}\sin(\theta_{\verb+t++1})\;+\;\sgn(\epsilon)\tau\eta\\
      \theta_{\verb+t++1}&=&\theta_{\verb+t+}\;+\;\sgn(\epsilon)J_{\verb+t+}\;\;\;\textrm{ mod } 2\pi
      \end{array}\right.
\end{equation}
If considered on the 2-torus, this map (\ref{classical})
has a fixed point at $J=0$ and $\theta=\theta_0$  if:
\begin{equation}
  \label{fixedpt}
%  J_0=2\pi m, m\in \zZ, \qquad
\sin\theta_0=-\sgn(\epsilon)\frac{\tau\eta}{\tilde{k}} \;,
\end{equation}
and this fixed point is stable if and only if:
\begin{equation}
  \label{stabfix}
  0\le \tilde{k}|\cos\theta_0| \le 4 \quad \textrm{ and } \quad
 \cos\theta_0=-\sgn(\epsilon) |\cos\theta_0| \;.
\end{equation}
Such stable fixed points give rise
 to stable islands immersed in a chaotic sea.  From (\ref{ItoJ}) it
 is immediately seen that, in physical momentum space, such islands travel  at
 constant velocity:
\begin{equation}
\label{constv}
  v=-\frac{\tau\eta}{\epsilon}\;.
\end{equation}
As islands trap some of the particle's wave packet, they give
rise to experimentally observable  quantum
accelerator modes \cite{oberthaler,Gil2006}. Because of such modes a quadratic growth of energy
is observed over significant time scales, both in the falling frame
and in the laboratory frame \cite{fishman2002}. However such quadratic growth eventually
comes to an end as the modes decay, due to tunneling out of the stable
islands \cite{sheinman}. Smaller accelerating  islands  may also exist, associated with
higher-order fixed points of map (\ref{classical}) \cite{italo2006,italo2008}, however we will restrict
ourselves to  the above described ones.

 \subsection{Long time asymptotics of the fidelity}

\begin{figure}[!ht]
\centering
\includegraphics[width=.5\textwidth]{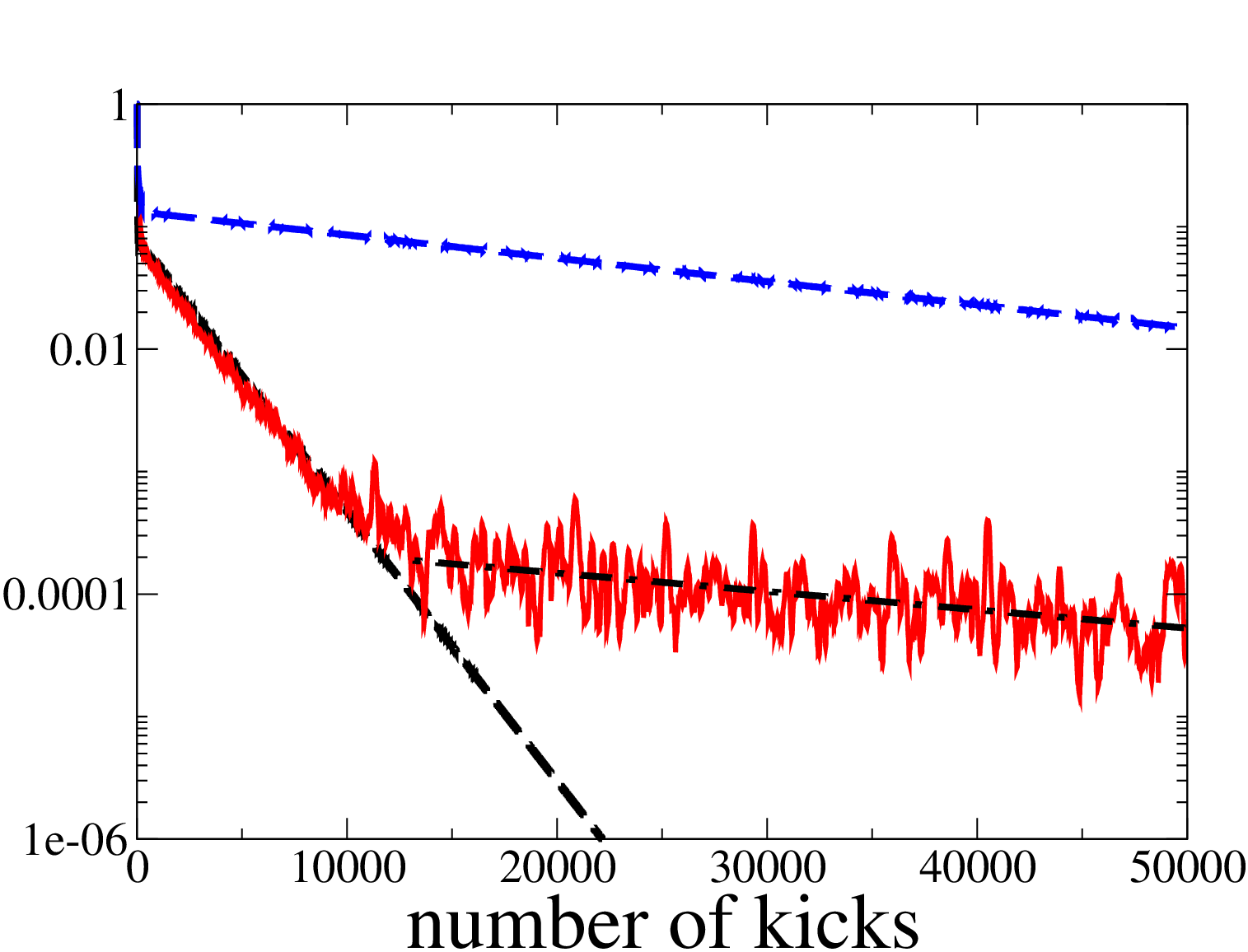}
\caption{(Color online) Comparison between the survival probability (black and blue dashed
  lines) and the fidelity (red full line). $\tau=5.86$, $\epsilon=\tau-2\pi$,
      $\eta=0.01579\tau$, $\beta=0.48984326$, $k_2=0.8\pi$,
      $k_1=0.7\pi$. The upper blue dashed line corresponds to $k_1$ and gives the
      tunneling rates $\Gamma_1 = 5.1 \times 10^{-4}$ whereas
      the lower black dashed line corresponds to $k_2$ and gives $\Gamma_2 =
      4.4 \times 10^{-5}$. The dot-dashed
      line corresponds to an exponential fit with a decay rate $\Gamma = 3.5 \times 10^{-5}$.}
\label{fid_vs_prob_decay}
\end{figure}

\begin{figure}[!ht]
\vspace{0.3cm}
\begin{minipage}{.32\linewidth}
  \begin{center}

    \includegraphics[width=\textwidth]{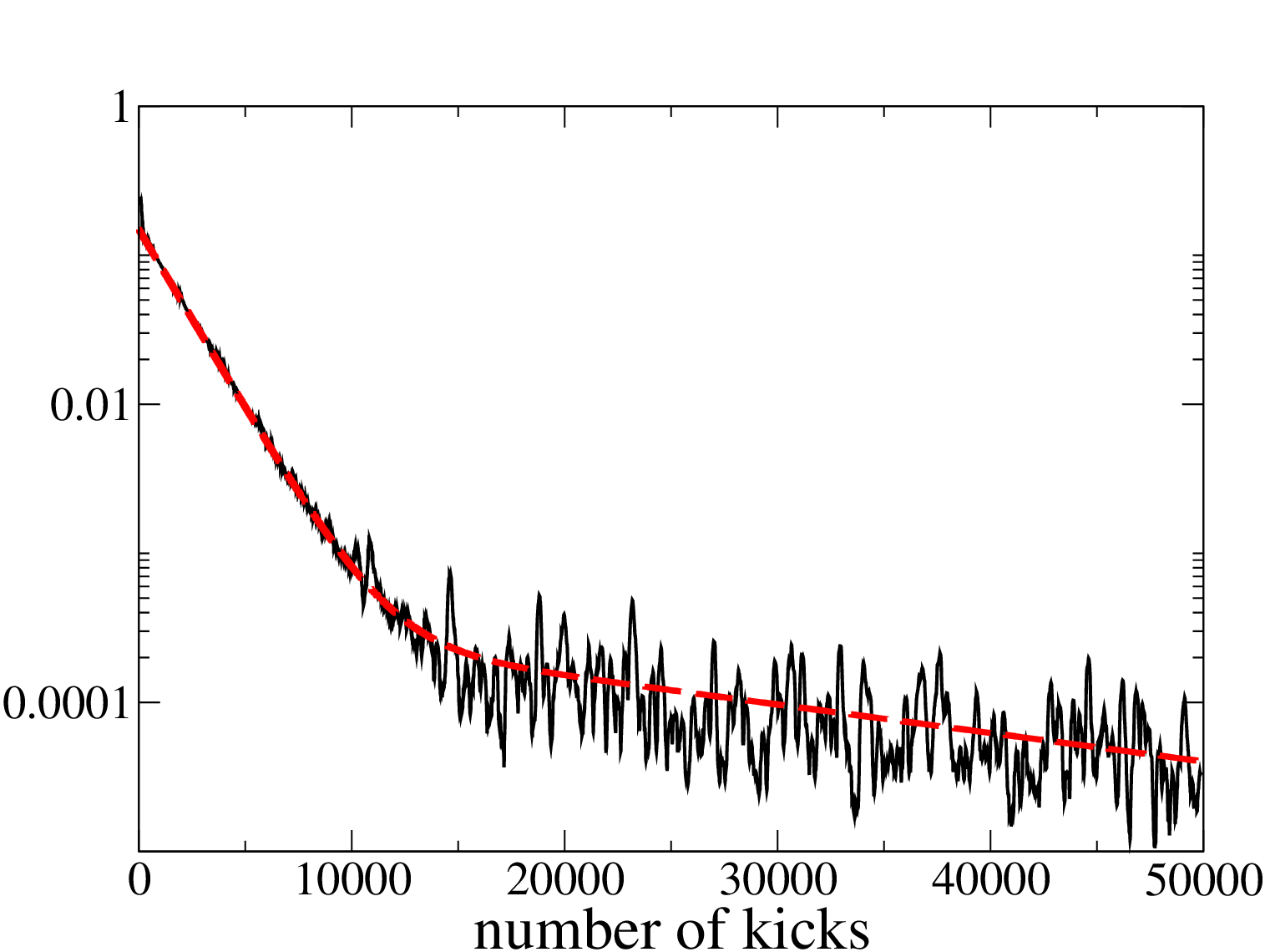}
\vspace{0.3cm}
    a)
  \end{center}
\end{minipage}
\begin{minipage}{.32\linewidth}
  \begin{center}

    \includegraphics[width=\textwidth]{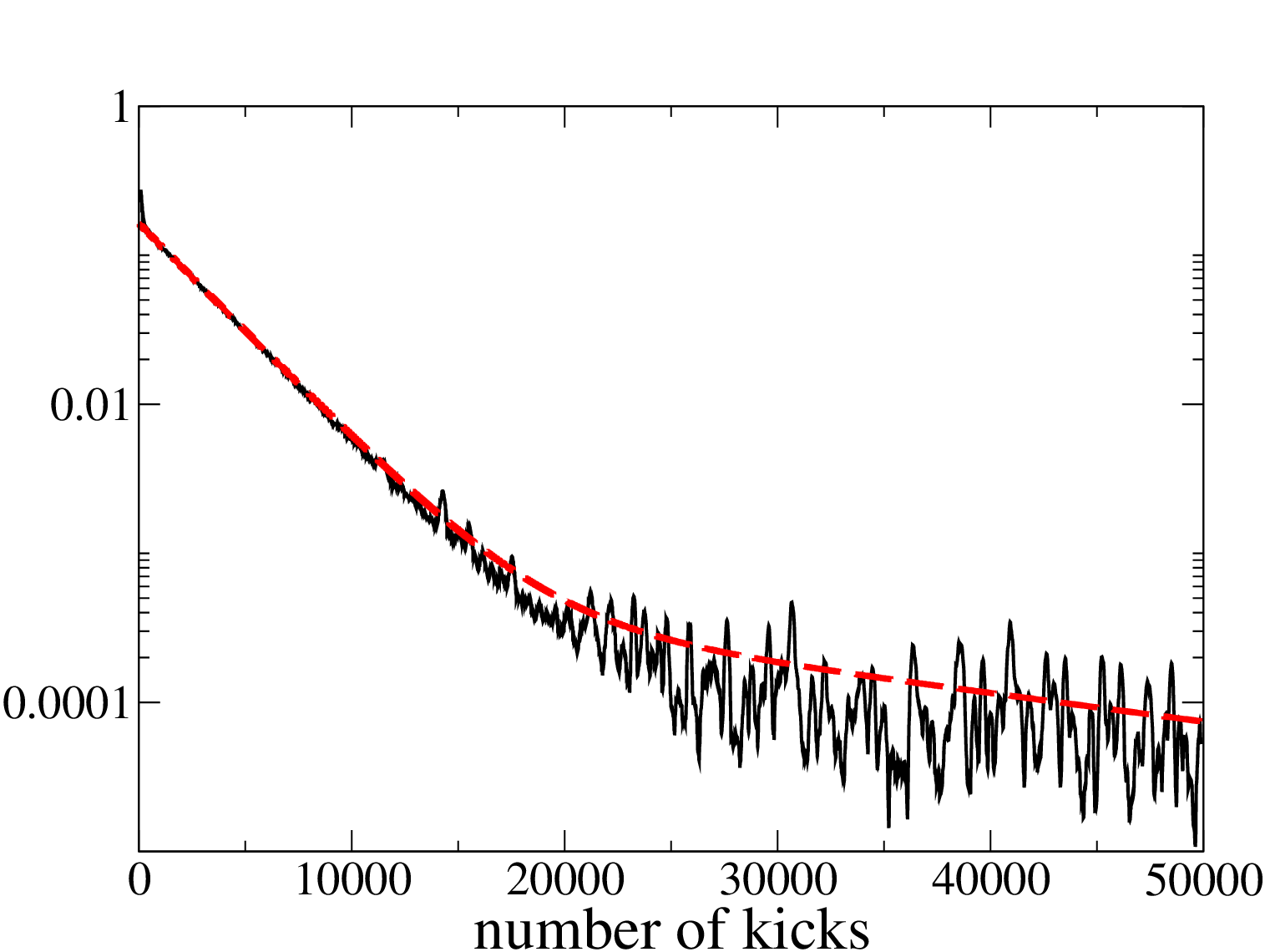}
\vspace{0.3cm}
    b)
  \end{center}
\end{minipage}
\begin{minipage}{.32\linewidth}
  \begin{center}
    \includegraphics[width=\textwidth]{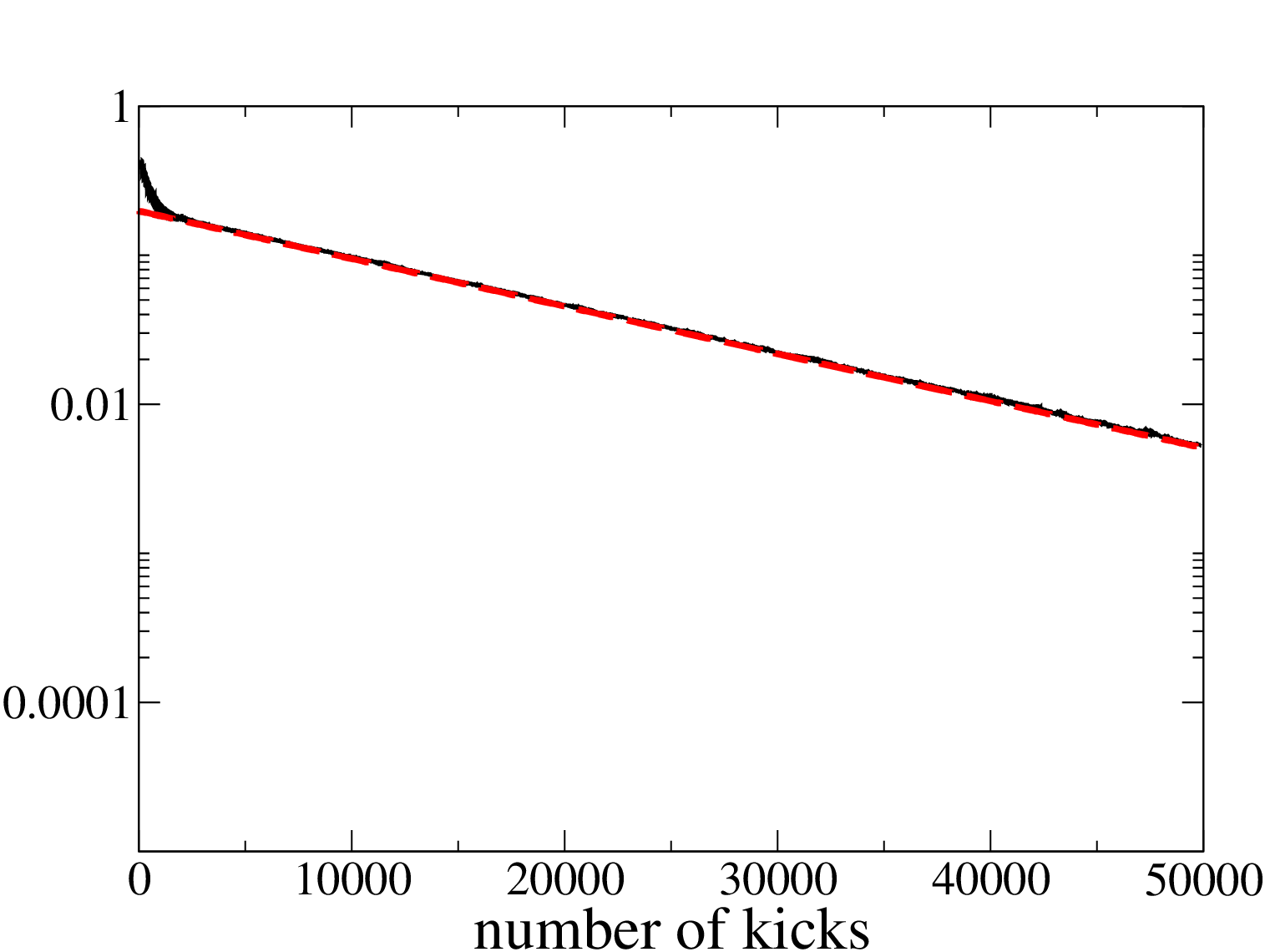}
\vspace{0.3cm}
    c)
  \end{center}
\end{minipage}
    \caption{(Color online) Comparison between the ansatz Eq.~(\ref{cldens}) (red dashed line) and the smoothed
      fidelity (black solid line) at $\tau=5.86$, $\epsilon=\tau-2\pi$,
      $\eta=0.01579\tau$, $\beta=0.48984326$, $k_2=0.8\pi$ and a)
      $k_1=0.7\pi$, b) $k_1=0.72\pi$, c) $k_1=0.83\pi$.}
\label{fidvstun1}
\end{figure}

\begin{figure}[!ht]
\vspace{0.3cm}
\begin{minipage}{.49\linewidth}
  \begin{center}

    \includegraphics[width=\textwidth]{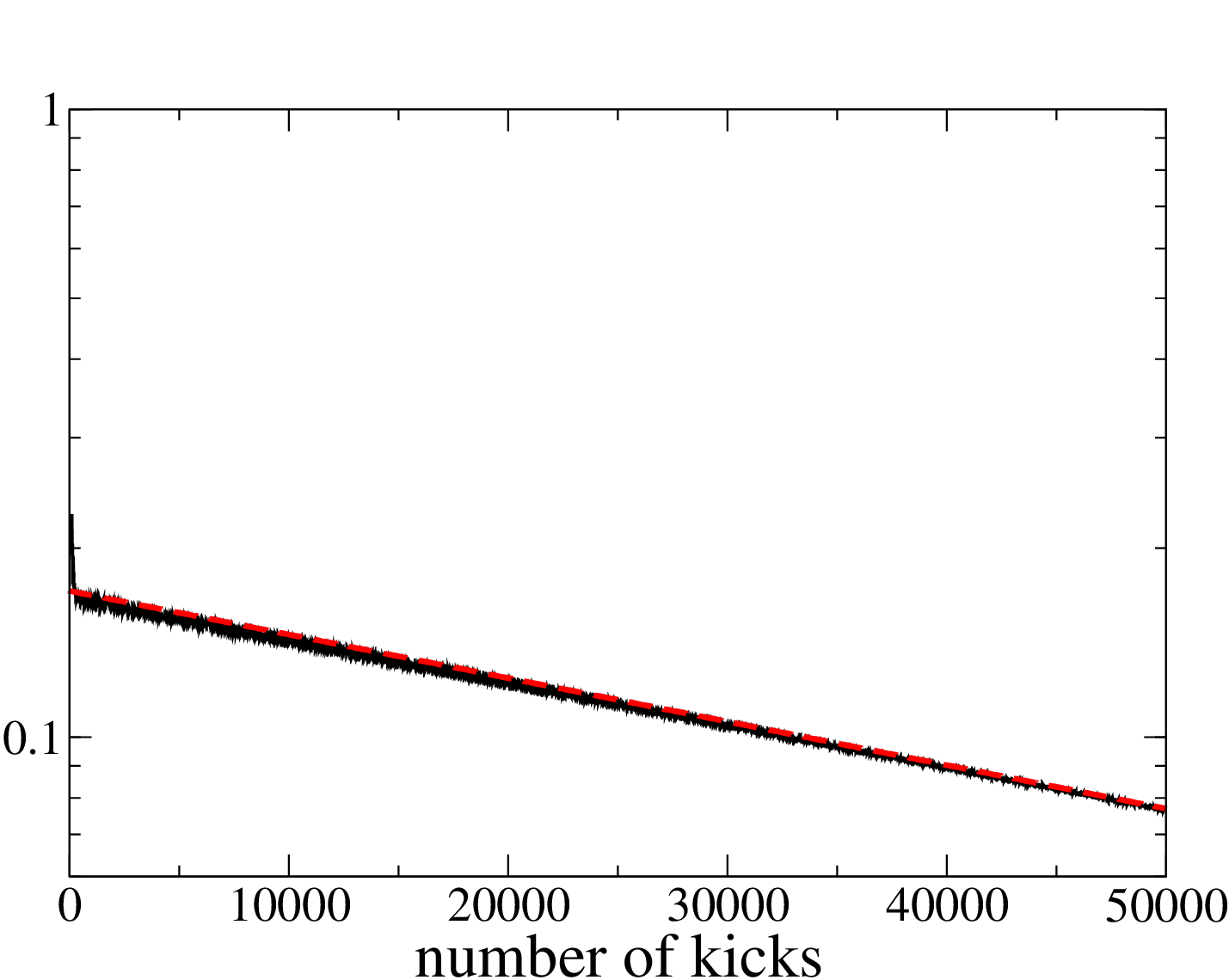}

    \vspace{0.3cm}
    a)
  \end{center}
\end{minipage}
\begin{minipage}{.49\linewidth}
  \begin{center}

    \includegraphics[width=\textwidth]{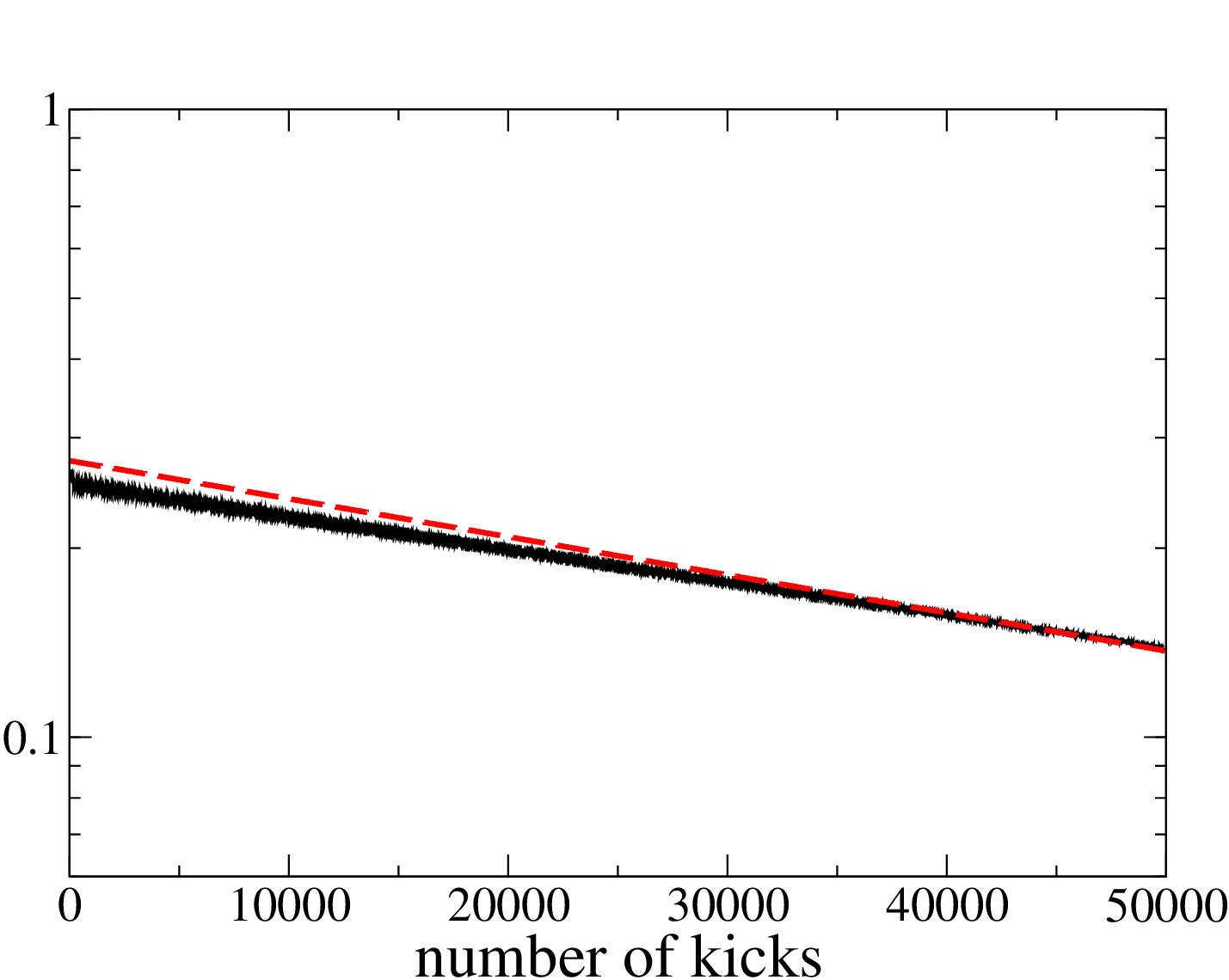}

    \vspace{0.3cm}
    b)
  \end{center}
\end{minipage}
    \caption{(Color online) Same as in Fig.~\ref{fidvstun1} at $\tau=6.6$, $\epsilon=\tau-2\pi$,
      $\eta=(\sqrt{5}-1)/20$, $\beta=0.123456789$,
      $k_2=2.5+\tau\eta$ and a) $k_1=2+\tau\eta$,  b) $k_1=3.5+\tau\eta$.}
\label{fidvstun2}
\end{figure}

\begin{figure}[!ht]
\vspace{0.3cm}
\begin{minipage}{.32\linewidth}
  \begin{center}

    \includegraphics[width=\textwidth]{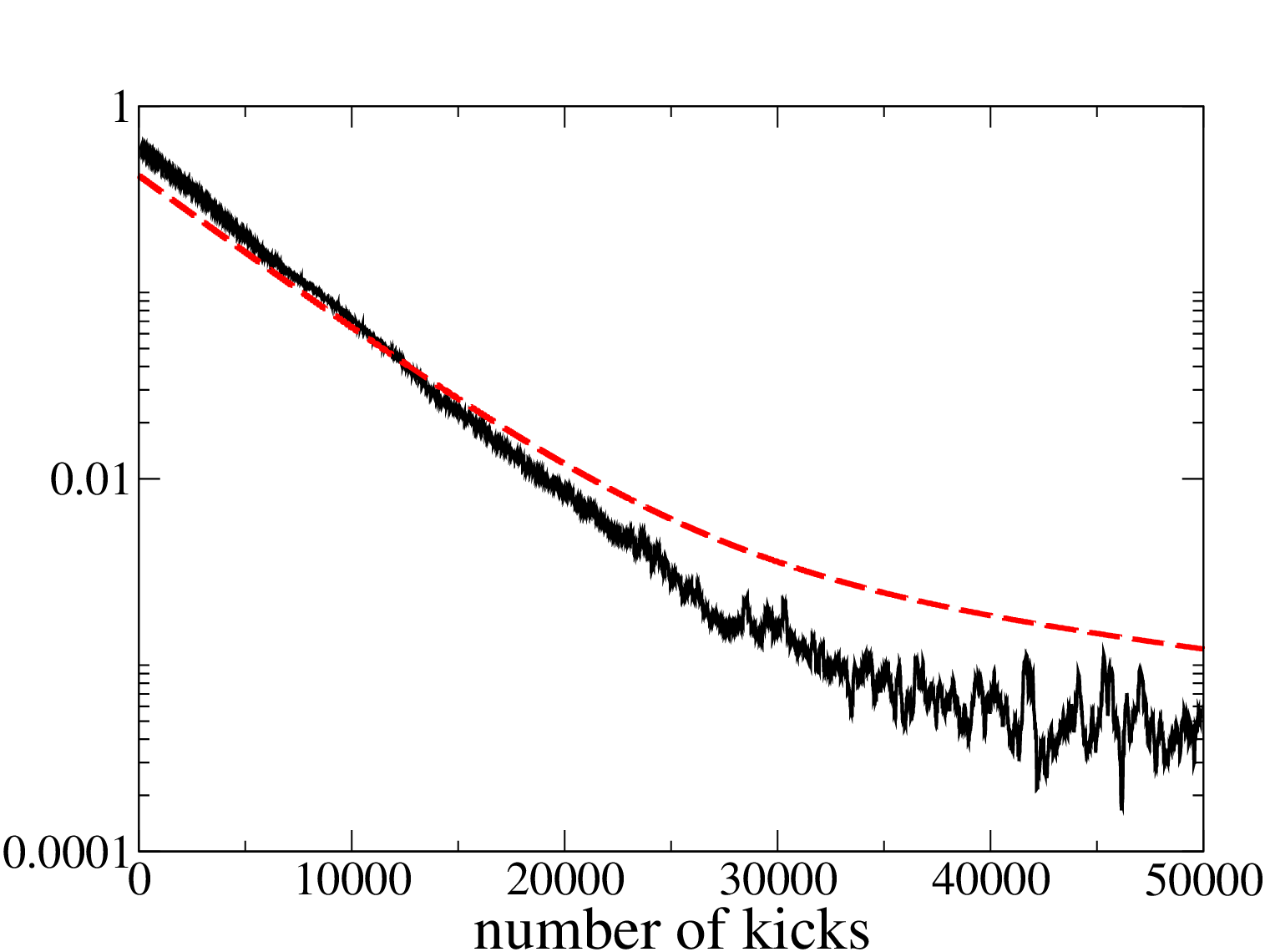}

\vspace{0.3cm}
    a)
  \end{center}
\end{minipage}
\begin{minipage}{.32\linewidth}
  \begin{center}

    \includegraphics[width=\textwidth]{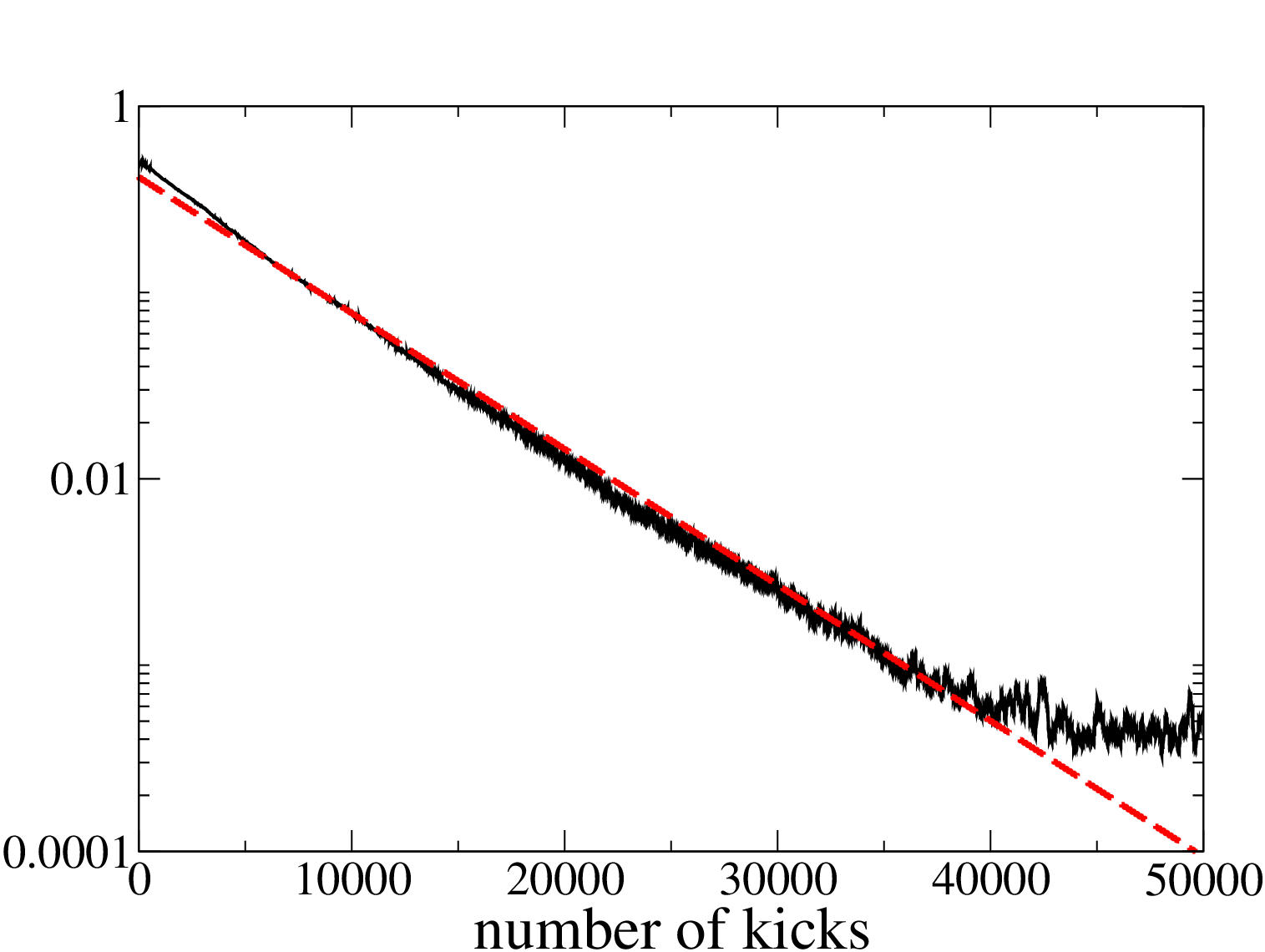}

\vspace{0.3cm}
    b)
  \end{center}
\end{minipage}
\begin{minipage}{.32\linewidth}
  \begin{center}
    \includegraphics[width=\textwidth]{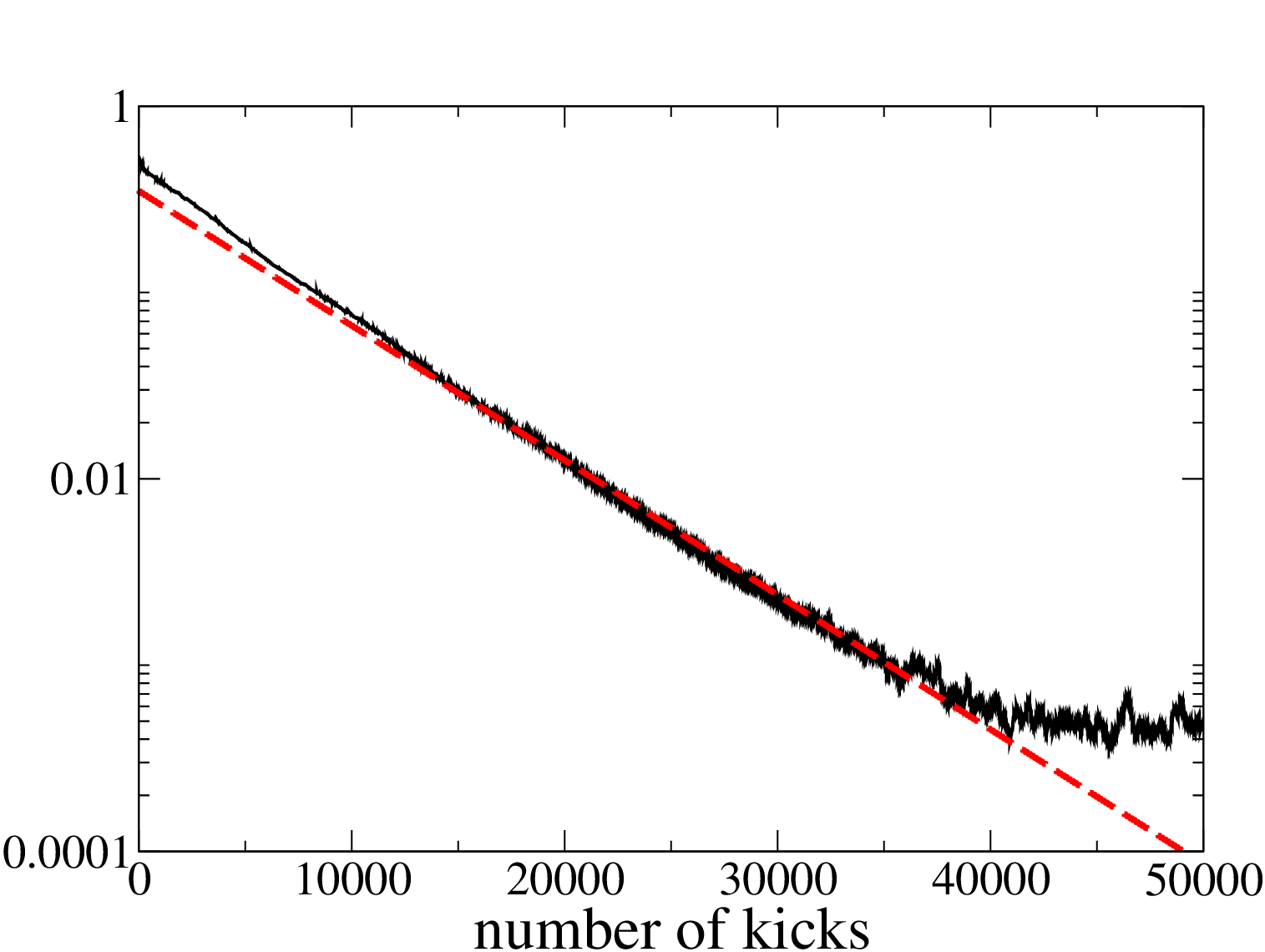}

\vspace{0.3cm}
    c)
  \end{center}
\end{minipage}
    \caption{(Color online) Same as Fig.~\ref{fidvstun1} at $\epsilon=-0.5$, $\tau=4\pi+\epsilon$,
      $\eta=0.001$, $\beta=0.123456789$, $k_2=(1.35+\tau\eta)/|\epsilon|$ and a)
      $k_1=(1+\tau\eta)/|\epsilon|$, b) $k_1=(2+\tau\eta)/|\epsilon|$, c)
      $k_1=(2.2+\tau\eta)/|\epsilon|$.%  The superimposed phasespaces are also
      % displayed in this last case.
    }
\label{fidvstun3}
\end{figure}

Typical numerical results illustrating the time dependence  of
fidelity (\ref{fid_def1rot}) are shown in
Figs.~\ref{fid_vs_prob_decay}--\ref{fidvstun3}.  In all those
simulations the parameters were chosen in  ranges where significant,
experimentally detectable  accelerator modes exist (see Appendix \ref{range_param}
for details). In general,    a very short initial transient is observed
(typically up to one or a few hundred kicks depending on parameters),
marked by a very quick drop.   A  clear, relatively long  exponential
decay follows. This is sometimes followed by yet another stage of
exponential decay, at a slower rate and with stronger
fluctuations. This general behavior is  qualitatively  understood as
follows.
 The initial sharp decay is due to the part of the initial wave packet
 that lies in the chaotic component of either of the two dynamics
 (defined by the two different kick strengths), and is rapidly
 carried away. The fidelity is thereafter  dominated by the parts of
 the wave packet which are trapped inside the islands.  For this
 reason, in order to bring this  stage of fidelity decay into full
 light, we choose our initial state in the form of a Gaussian state
 mostly located inside one island.  The islands which correspond to
 the $k_1$ and to the $k_2$ dynamics are slightly displaced with
 respect to each other, however they travel in momentum space with
 the same velocity (\ref{constv}). Therefore, the mismatch  between
 the $k_1$ and the $k_2$ dynamics  is, in $\epsilon$-classical terms,
 mostly  produced  by (i) different structures inside the islands and
 (ii) the decay from the islands into the chaotic sea due to dynamical
 tunneling. Concerning (i), the different rotational frequencies in the
 two islands are expected to produce quasi-periodic oscillations of
 the fidelity \cite{abb,ott_fishman2011}, so (ii)  should be the main mechanism responsible of
 the mean fidelity decay.
 This leads to the following crude description. The main contribution
 to  fidelity  comes from the part of each factor in the scalar product
 in (\ref{fid_def1rot}) which is trapped in the respective travelling
 island. Hence the decay of fidelity is determined by the decay of each
 part,  which is in turn  determined by its respective tunneling rate
 into the  chaotic sea,  which we will denote by $\Gamma_{i}$ for
 $i=1,2$. Then a  simple, self-explanatory  ansatz for the asymptotic
 decay of fidelity is
\begin{equation}
  \label{cldens}
  F(\verb+t+)\;\propto\;{\mu({\cal A}_1\backslash{\cal A}_2)\; e^{-\Gamma_1 \verb+t+} +
    \mu({\cal A}_2\backslash{\cal A}_1)\; e^{-\Gamma_2 \verb+t+} + \mu({\cal
        A}_2\cap{\cal A}_1)\;e^{-(\Gamma_1+\Gamma_2) \verb+t+}} \ ,
\end{equation}
where $\mu$  is the classical invariant measure of phase space sets,  ${\cal A}_i$ is the island 
around the fixed point associated with $k_i$.  This ansatz was found to satisfactorily reproduce the actual
decay of fidelity in our numerical checks.   In our simulations, the
phase-space areas appearing in (\ref{cldens})  were estimated as described
in appendix~\ref{algo}.  Quantum decay rates were found as follows:
for both  the $k_1$ and the $k_2$ dynamics we   numerically computed the
probability  in a momentum  range centered on the accelerator mode. This
measures the amount of the initial probability,
which  travels within the accelerator mode. This quantity is called the
survival probability and shown in Fig.~\ref{fid_vs_prob_decay}. Fitting the long time decay of this
probability with an exponential function gives us an estimate of the
tunneling rate, see an example in Fig.~\ref{fid_vs_prob_decay}. In our
numerical simulations we take the following momentum range:
\begin{equation}
  \label{windows_mom}
  [n(\verb+t+)-15\;,\;n(\verb+t+)+15], \qquad n(\verb+t+)=n_0+v\verb+t+ \ ,
\end{equation}
where $n_0$ and $v$ are given respectively by (\ref{param_init}) and
(\ref{constv}). We checked that the peak traveling
ballistically has a width less than $30$ (in two photonic recoil units for the
experiment, see, e.g., \cite{review}) with our choice of parameters.

We note that the right hand side in Eq.~(\ref{cldens})  is defined up to a
proportionality factor. Moreover, it crucially depends on $k_1$ and $k_2$,
because island sizes and tunneling rates  vary when the kicking strength is
changed. In our simulations the proportionality factor was chosen such as  to
fit  the earlier regime of  exponential decay
(approximately between 100 and $10^4$ kicks in Fig.~\ref{fid_vs_prob_decay} for instance).
 In our numerical
computations the initial state is given by (\ref{psi_init_2}) with a width $\sigma^2=0.25$. %_{\rm \epsilon}
%given in units of the rescaled $\epsilon$-classical momentum, see Eq.~\eqref{scaledmom}.
 As the fidelity   is a wildly oscillating
function, it is averaged over 200 kicks in order to  clearly expose the
mean long time behavior. Results are shown in the
Figs.~\ref{fidvstun1},\ref{fidvstun2}, and \ref{fidvstun3} for different sets
of parameters and will be discussed in the following.
%------------------------   set1
%--------------------------- SET2
%--------------------------- SET3

\begin{figure}[!ht]
\vspace{0.3cm}
% \begin{minipage}{.49\linewidth}
%   \begin{center}

% %    \includegraphics[width=\textwidth]{phasespace_set1_k1_0.7pi_k2_0.8pi.eps}
%     \includegraphics[width=\textwidth]{fig7a.eps}

%     \vspace{0.3cm}
%     a)
%   \end{center}
% \end{minipage}
% \begin{minipage}{.49\linewidth}
%   \begin{center}

% %    \includegraphics[width=\textwidth]{phasespace_set3_k1_4.424_k2_2.724.eps}
%     \includegraphics[width=\textwidth]{fig7b.eps}

%     \vspace{0.3cm}
%     b)
%   \end{center}
% \end{minipage}
    % \caption{(Color online) Superimposed phase spaces corresponding to $k_1$
    %   (black dots) and
    %   $k_2$ (red $+$). a)  $\tau=5.86$, $\epsilon=\tau-2\pi$, $\eta=0.01579\tau$, $k_1=0.7\pi$,
    %   $k_2=0.8\pi$.  b) $\epsilon=-0.5$, $\tau=4\pi+\epsilon$,  $\eta=0.001$,
    %   $k_1=(2.2+\tau\eta)/|\epsilon|$, $k_2=(1.35+\tau\eta)/|\epsilon| $.}
\includegraphics[width=\textwidth]{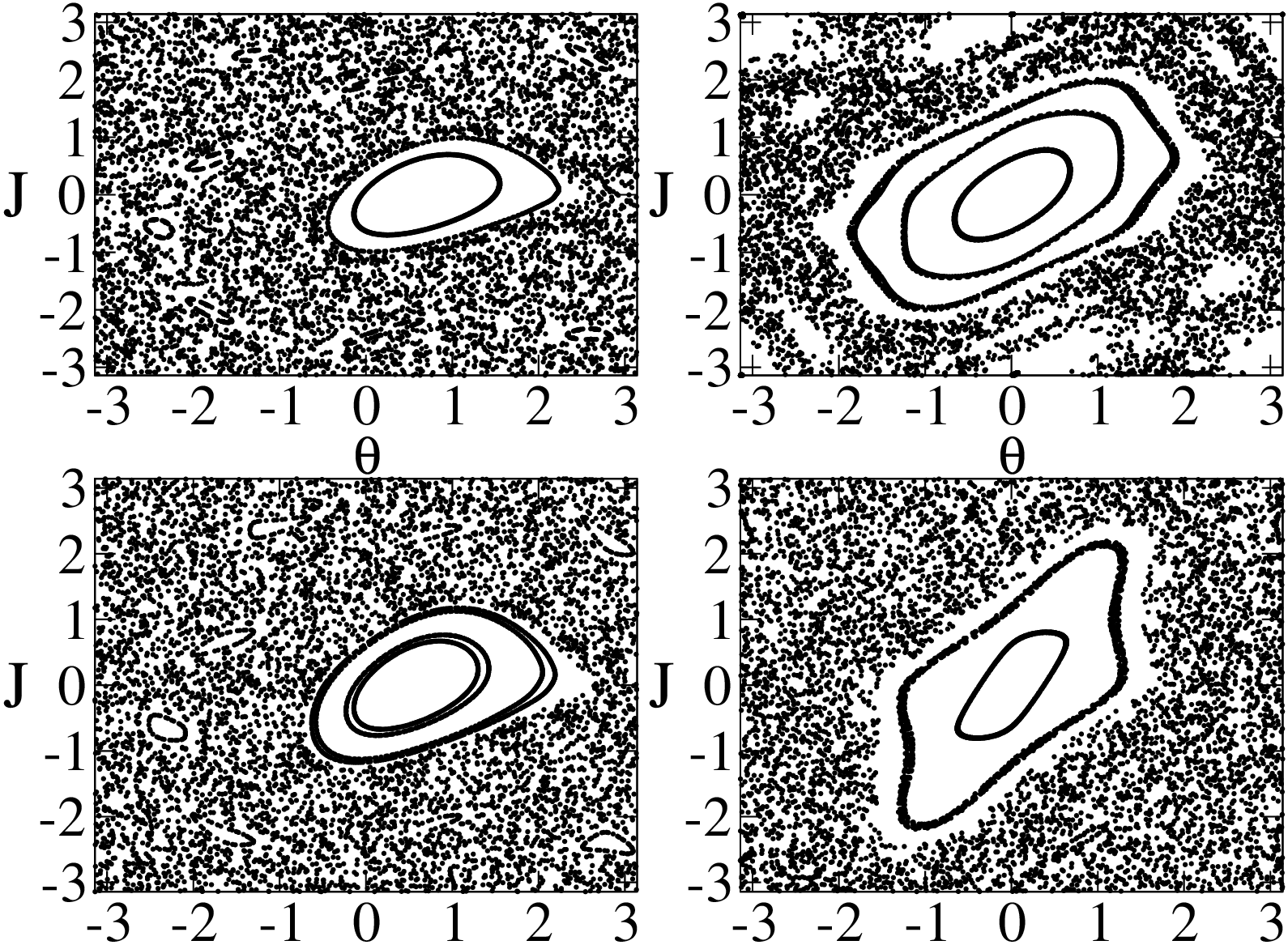}
    \caption{Phase spaces corresponding to two sets of parameters. Left
      column: $\tau=5.86$, $\epsilon=\tau-2\pi$, $\eta=0.01579\tau$. Top left:
      $k=0.7\pi$. Bottom left: $k=0.8\pi$. Right column: $\epsilon=-0.5$,
      $\tau=4\pi+\epsilon$, $\eta=0.001$. Top right:
      $k=(1.35+\tau\eta)/|\epsilon|$. Bottom right: $k=(2.2+\tau\eta)/|\epsilon| $.}
\label{bothphsp}
\end{figure}

The mean behavior of the fidelity for long times is well
reproduced by the ansatz (\ref{cldens}) when $\Gamma_1$ and
$\Gamma_2$ are quite different: then the fidelity  shows successively two different
decay regimes, which are  well reproduced by the pseudo-classical ansatz
(\ref{cldens}), see Figs.~\ref{fidvstun1}a ($\Gamma_1\simeq 10\Gamma_2$) and
\ref{fidvstun1}b ($\Gamma_1\simeq 5\Gamma_2$). On the contrary, when $\Gamma_1$
and $\Gamma_2$ are close to each other, one can see only one decay, which is
still  well reproduced  by (\ref{cldens}), see Fig.~\ref{fidvstun1}
($\Gamma_1\simeq \Gamma_2$) and Fig.~\ref{fidvstun2} ($\Gamma_1\simeq 3\Gamma_2$).
For sake of comparison we are showing the same plots for $l=2$ in Fig.~\ref{fidvstun3}.
The quantum resonance is then $\tau=4\pi$. It can be seen that the agreement is not so
good in the latter case. One reason for this  may be that higher order $\epsilon-$classical phase space structures
 have a larger area hence may play a more
important role, making estimates of the various areas in (\ref{cldens}) more problematic.
Both $\epsilon-$classical phase spaces corresponding
to $k_1$ and $k_2$ are displayed in Fig.~\ref{bothphsp} corresponding to
Fig.~\ref{fidvstun1}a and Fig.~\ref{fidvstun3}c. It is clear that the
shape and the overlap between the two islands are qualitatively different in
these two situations.

\section{Conclusion}

We have presented a theoretical analysis of the temporal dependence of fidelity for the
quantum kicked rotor subject to an additional gravity field. The two major results concern
the dynamics of this system at principal quantum resonances, i.e., at kicking periods $\tau=2\pi l$ ($l$ integer),
and close to these resonances. In the former case, we arrive at analytical estimates for the decay of fidelity
which is highly sensitive to the arithmetic properties of the gravity parameter.
Close to a resonance we have used the $\epsilon-$semiclassical method in order to describe
the long time asymptotics of the fidelity.
The ansatz (\ref{cldens}) based on semiclassical densities gives a good description of the long time
behavior of the quantum fidelity for both similar and different tunneling rates of the two compared non-dispersive wave packets
\cite{buch} centered at the accelerator mode islands in phase space \cite{sheinman}.
This result highlights once more the utility of $\epsilon-$classics in describing the quantum evolution of the kicked rotor
and its variants.

\acknowledgments

It is a pleasure to thank D. Ullmo, J. Marklof, P. Schlagheck and G. S. Summy for helpful
discussions. R.D. and S.W. acknowledge financial support from the DFG through FOR760, the Helmholtz
Alliance Program of the Helmholtz Association (contract HA-216
Extremes of Density and Temperature: Cosmic Matter in the
Laboratory) and within the framework of the Excellence Initiative
through the Heidelberg Graduate School of Fundamental Physics (grant
number GSC 129/1), the Frontier Innovation Fund and the Global
Networks Mobility Measures during R.D.'s stay at the University of Heidelberg
when this work was done.

\appendix

\section{Relevant range for the parameters.}
\label{range_param}

It was observed numerically, when plotting the $\epsilon-$classical phase portrait, that
higher order nonlinear resonances play a bigger role when $l$ is increased.
For this reason we restrict in this paper mainly to $l=1$.
In the experiments, see
e.g. \cite{exp}, $k$ typically runs
from $0.3\pi\sim0.94$ to $1.5\pi\sim4.71$. In order to see an
accelerator mode one needs a stable fixed point of the classical
map. Following (\ref{fixedpt}), for a given $k$ one has a fixed
point when $\tilde{k}\ge \tau \eta$. Taking into account experimental
constraints leads us to choose $\epsilon$ such that:
\begin{equation}
  \label{range_eps}
  0.4 < |\epsilon|< 1 \,.
\end{equation}
For the quantum fidelity the initial state is chosen to be a Gaussian state:
\begin{equation}
  \label{psi_init_2}
  \left<n|\psi_\beta(0)\right>=\frac{e^{-(n-n_0)^2/4\sigma^2+\ic n\theta_0}}{(2\pi\sigma^2)^{1/4}}\ ,
\end{equation}
with
\begin{equation}
  \label{param_init}
  \sin\theta_0=-\frac{\tau\eta}{\epsilon k}, \quad n_0=\frac{2\pi\
    m}{|\epsilon|}-\frac{1}{\epsilon}\left[\pi l +
    \tau\left(\beta+\frac{\eta}{2}\right)\right] \,,
\end{equation}
where $m$ is an arbitrary integer. This initial state is centered on an accelerator mode
for which $J=0$. This mode is travelling (in momentum space) at the speed
(\ref{constv}), see e.g. \cite{fishman2002}. The fidelity is computed by
applying successively the operators (\ref{Flqt}) for two different
values of the kicking strength: $k_1$ and $k_2$. For each set of parameters we keep $k_2$ fixed and vary
$k_1$. Our initial state is always chosen such as to follow the accelerator
mode attached to $k_1$: once $\tau$ (and $\epsilon$), $\eta$ and $\beta$ are
fixed, the initial state is centered in momentum space around $n_0$ defined by (\ref{param_init}).

\section{Estimating the size of the  islands in the $\epsilon-$classical
  phase space}
\label{algo}
\begin{figure}[!ht]
\centering
\vspace{1cm}
\includegraphics[width=.5\textwidth]{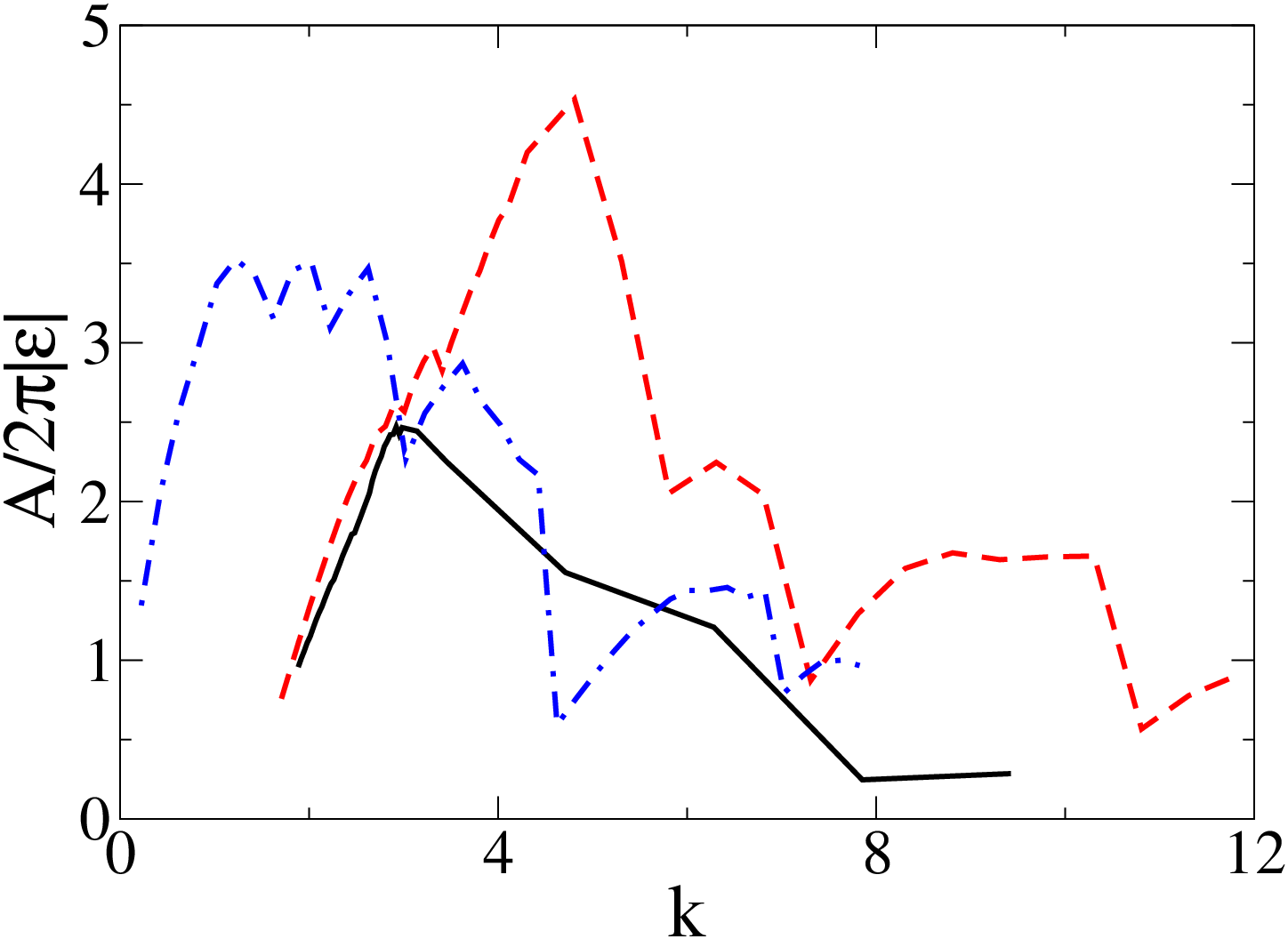}
\caption{(Color online) Area of the stable island in the $\epsilon-$classical phase
  space as a function of the kicking strength for three sets of parameters. Black
  full line: $\tau=5.86, \epsilon=\tau-2\pi\sim -0.42, \eta=0.01579\tau$. Red dashed
  line: $\tau=6.6, \epsilon=\tau-2\pi\sim 0.32, \eta=\phi/10$. Blue dash-dotted line:
  $\epsilon=-0.5, \tau=4\pi+\epsilon, \eta=0.001$.}
\label{area}
\end{figure}

First we choose a set of
parameters $\tau$ and $\eta$ for which we can have
accelerator modes. Then we vary $k$ within the range of existence of
these modes. One way to see this range is to compute numerically the
area of the stable island in the $\epsilon-$classical phase space, see
Fig.~\ref{area}.
The area follows a bell shape as a function of $k$. In
Fig.~\ref{area} some jumps are also visible (see e.g. for the blue dash-dotted line,
between $k\simeq4.424$ and $k\simeq4.624)$. We believe that this is
due to the lack of precision when determining the island boundary and/or the
breaking of outermost tori and their remnants (cantori). %, see Fig.~\ref{check_area}.
The size of the stable island in the $\epsilon-$classical phase space is
computed by starting a fairly small number of trajectories
outside the island. These are typically iterated for a long time ($10^8$ kicks).
Then we move to polar coordinates
$(\varphi,I)$ centered at the fixed point under interest \cite{peter}. 
The boundary of the island is then determined by the curve $I(\varphi)$ defined in
the following way. Using a grid of thickness $\delta\varphi$ along the $\varphi$ axis,
the boundary $I_i=I(\varphi_i)$ is given by:
\begin{equation}
  I_i=\min \left\{ I_j, (\varphi_j,I_j) \textrm{ iterated points and }
    \varphi_i\le\varphi_j\le\varphi_i+\delta\varphi\right\}
\end{equation}
The size of the island is given by the area under this curve. Numerically it
is computed via a Riemannian sum.\\
The measures of the different sets in (\ref{cldens}) are computed by propagating a
cloud of $10^4$ classical points. The initial points are distributed following
normal distributions with mean and width $(\theta_0,\sigma_\theta)$ in the
$\theta$ direction and $(J_0,\sigma_J)$ in the $J$ direction. 
The areas in (\ref{cldens}) reached stationary values after ca $500$
kicks. The measures needed in (\ref{cldens}) are simply given by the number of
points sitting on one or both of the stable islands associated to $k_1$ and
$k_2$, respectively.

\end{document}